\documentclass[letterpaper, 10 pt, conference]{ieeeconf} 

\IEEEoverridecommandlockouts                             
\usepackage{cite}
\usepackage{comment}
\usepackage{amsmath,amssymb,amsfonts}
\usepackage{graphicx}
\usepackage{textcomp}
\usepackage{balance}
\usepackage{tabularx}
\usepackage{makecell}
\usepackage{multirow}
\usepackage{algorithm}
\usepackage{algpseudocode}
\usepackage{url}
\usepackage{hyperref}
\usepackage{subcaption} 
\usepackage[table,xcdraw]{xcolor}

\def\BibTeX{{\rm B\kern-.05em{\sc i\kern-.025em b}\kern-.08em
    T\kern-.1667em\lower.7ex\hbox{E}\kern-.125emX}}

\title{\LARGE \bf
OceanLens: An Adaptive Backscatter and Edge Correction using Deep Learning Model for Enhanced Underwater Imaging
}

\author{Rajini Makam$^{1*}$, Dhatri Shankari T M$^{2}$, Sharanya Patil$^{3}$, and Suresh Sundram$^{1}$ 
\thanks{$^{*}$Corresponding Author} %
\thanks{$^{1}$Rajini Makam \& Suresh Sundaram are with the Department of Aerospace Engineering, Indian Institute of Science, Bangalore, India.
        {\tt\small \{\{rajinimakam,vssuresh\}@iisc.ac.in\}}}%
\thanks{$^{2}$ Dhatri Shankari T M is with the Dept. of Elect. and Comm. Engn., PES University, Bangalore, India. {\tt\small \{dhatristm2003@gmail.com\}}}%
\thanks{$^{3}$ Sharanya Patil is with the Dept. of Computer Science Engn., PES University, Bangalore, India. {\tt\small \{sharanyapatil23@gmail.com\}}}%
}

\begin{document}

\maketitle
\thispagestyle{empty}
\pagestyle{empty}

\begin{abstract}
 
Underwater environments pose significant challenges due to the selective absorption and scattering of light by water, which affects image clarity, contrast, and color fidelity. To overcome these, we introduce OceanLens, a method that models underwater image physics—encompassing both backscatter and attenuation—using neural networks. Our model incorporates adaptive backscatter and edge correction losses, specifically Sobel and LoG losses, to manage image variance and luminance, resulting in clearer and more accurate outputs. Additionally, we demonstrate the relevance of pre-trained monocular depth estimation models for generating underwater depth maps. Our evaluation compares the performance of various loss functions against state-of-the-art methods using the SeeThru dataset, revealing significant improvements. Specifically, we observe an average of 65\% reduction in Grayscale Patch Mean Angular Error (GPMAE) and a 60\% increase in the Underwater Image Quality Metric (UIQM) compared to the SeeThru and DeepSeeColor methods. Further, the results were improved with additional convolution layers that capture subtle image details more effectively with OceanLens. This architecture is validated on the UIEB dataset, with model performance assessed using Peak Signal-to-Noise Ratio (PSNR) and Structural Similarity Index Measure (SSIM) metrics. OceanLens with multiple convolutional layers achieves up to 12-15\% improvement in the SSIM. \url{https://github.com/AIRLabIISc/OceanLens}
\end{abstract}

\section{Introduction}

Exploring the depths of the ocean through imaging presents various challenges and obstacles due to the optical characteristics of underwater surroundings. These challenges stem from water's ability to absorb and scatter light waves selectively based on their wavelength preferences \cite {shuang2024}. The presence of these factors can significantly abate the clarity of images by diminishing the contrast \cite{akkaynak2017} and distorting colors while introducing noise that can obscure details and make accurate visual understanding difficult, consequently resulting in underwater images lacking the sharpness and detail compared to those found in their aerial counterparts.

Scattering, especially by suspended particles and microorganisms  \cite{Wang2019}, disperses light on a wider scale—reducing image sharpness and loss of clear detail. The other effect is backscattering \cite{akkaynak2018} where light is scattered straight back towards the camera, which contributes to increased noise and decreased contrast. Spectrally Selective Light Attenuation leads to a diminishing signal-to-noise ratio making it difficult to capture detailed images specifically in murkier waters complicating the capture of finer details \cite{jamieson2020}. The optical properties of natural water bodies differ drastically from that of the atmosphere. Coastal ports of the ocean are brownish and muddy but offshore blue-green hues dominate other colors further reducing color fidelity \cite{liang2023}.

Underwater images, much like terrestrial ones, often suffer from a dense, fog-like distortion caused by insufficient ambient light \cite{akkaynak2019}. Traditional enhancement techniques often fall short in effectively improving the visual quality of these images \cite{zhu2020}. However, the recent surge in available underwater datasets has opened new avenues for applying deep learning techniques to enhance underwater imagery. These advanced methods can significantly improve image quality while also reducing computational demands \cite{Jamieson2023}. Despite access to an abundance of large image datasets, the oceanic domain has yet to fully exploit the potential of deep learning techniques, especially those that are computationally efficient. While recent advancements have been promising, current models still face challenges, particularly in accurately preserving edges, adapting to varying lighting conditions, and not only maintaining color fidelity but also achieving the full range of color accuracy in underwater environments.

In this paper, we propose a novel architecture, OceanLens, inspired by the \cite{Jamieson2023} framework. The model employs edge loss functions—Laplacian of Gaussian (LoG) and Sobel—for enhanced edge detection and detail capture. Additionally, we incorporate an adaptive backscatter loss to manage varying lighting conditions. Our approach also explores the use of monocular depth estimation models, such as MonoDepth2 \cite{godard2019} and Depth-Anything-V2-Large \cite{yang2024v2}, demonstrating that these models can deliver performance comparable to traditional Structure-from-Motion (SfM) techniques for underwater image enhancement. Our results are compared against SeeThru and DeepSeeColor using the GPMAE metric, showing significant improvements and enhanced UIQM, indicating better overall image quality. Additionally, we enhance the OceanLens architecture by adding additional convolution layers, increasing the network’s capacity to capture subtle image features. This architecture is validated with the UIEB dataset. Since UIEB dataset has reference images, we show our model's efficacy with PSNR and SSIM metrics. 

The paper is structured as follows: Section \ref{s2} provides an overview of the existing work in the field of underwater image enhancement. Section \ref{s3} presents the proposed OceanLens methodology. Results obtained from the OceanLens method are discussed in Section \ref{s4}. Section \ref{s5} provides the conclusion of the work.
 
\section{Related Work} \label{s2}

Underwater image enhancement has significantly advanced in recent years, especially when new comprehensive datasets and novel algorithms appeared. Due to the unique challenges posed by aquatic environments, underwater image enhancement has attracted widespread attention.

\subsection{Pre-Processing Methods}

Commonly used techniques for underwater image enhancement include dehazing \cite{berman2017, han2018}, white balance correction \cite{sanila2019}, gamma correction \cite{hong2019}, and histogram equalization \cite{hitam2013}. Gamma correction adjusts the image intensity to compensate for the reduced contrast caused by light absorption in water, helping to restore visual clarity \cite{hong2019}. The Dark Channel Prior (DCP) \cite{he2010} is a well-known dehazing method that mitigates scattering effects by leveraging the physical properties of light propagation in water. White balancing \cite{li2018} tackles the issue of color cast by adjusting color channels to improve the natural appearance of underwater images. 

\subsection{Image Enhancement Methods}

The performance of computer vision algorithms for image classification and scene reconstruction is significantly hindered by color distortions, affecting both deep neural networks and traditional feature detection methods \cite{zhou2014}, \cite{de2021}. Fusion-based models have recently gained traction, combining feature extraction with color and white balance correction \cite{zhou2022}, and hybrid fusion techniques \cite{an2024} to address underwater photography's unique challenges. Traditional methods often struggle with the complexity of underwater imagery.

Deep learning models, particularly convolution neural networks (CNNs), have shown promise in overcoming these limitations by being trained on large underwater datasets to perform robust color correction \cite{peng2022}. Physics-based enhancement techniques further contribute by simulating how light interacts with water molecules, allowing for the approximation of image formation parameters to accurately reverse color shifts, enhance contrast, and reduce noise \cite{chandrasekar2024}, \cite{Pham2023}.

The work in \cite{li2019} introduced the UIEB dataset, a large-scale collection of 950 underwater images, and proposed Water-Net, a CNN using learned confidence maps to enhance image quality. However, it faces limitations in maintaining color reconstruction consistency under varying illumination conditions, making it less reliable than physics-based methods. Other approaches use generative adversarial networks (GANs) for paired and unpaired training data enhancement \cite{islam2020}, or generate synthetic underwater images from in-air scatter images and depth maps \cite{li2017}. These methodologies illustrate the ongoing efforts to refine underwater image enhancement, each with its strengths and challenges.

\subsection{Use of Range-Maps for Image Reconstruction}

Originally, the underwater image formation model relied on using multiple color chart calibration targets to validate color correction and image restoration accuracy \cite{akkaynak2018}. Depth maps have since emerged as valuable tools for enhancing underwater images by providing crucial 3D information that aids in improving various aspects of image quality \cite{yu2023}. Depth-driven enhancement techniques utilize this depth information to adjust contrast and correct colors, effectively highlighting regions of interest based on their depth \cite{chen2023}. Some methods even employ stereo imaging for depth estimation \cite{berman2020}. Monocular depth estimation has proven crucial for enabling real-time 3D perception of underwater environments, as shown by Bryson et al. \cite{bryson2016}, who demonstrated that accurate distance maps could generate high-resolution images from low-resolution underwater scenes.

Akkaynak et al. \cite{akkaynak2019} introduced a method that significantly enhances image clarity by leveraging the physics of light propagation in water. By estimating range-dependent attenuation coefficients for both direct and backscattered signals, their approach employs structure-from-motion (SfM) \cite{hu2023} to generate range maps, often used to measure the structural complexity of underwater reefs. While some earlier works focused on high-quality offline color reconstruction \cite{berman2017}, \cite{bryson2016}, Jamieson et al. \cite{jamieson2020} proposed an adaptive color correction algorithm using range maps generated from stereo images to tackle issues like light attenuation and color distortions, achieving real-time processing.

Other approaches have incorporated depth map optimization and background light estimation \cite{liu2022}, while some estimate illumination through depth mapping to automatically adjust the global color distribution of images \cite{Jamieson2023}, \cite{chandrasekar2024}, \cite{zhou2021}. These methods have shown effectiveness in enhancing specific aspects like contrast and depth map generation.

To the best of the authors' knowledge, existing models can enhance aspects like contrast, saturation, illuminance, and variance, but they struggle with edge preservation, adapting to varying lighting conditions, and maintaining color accuracy in underwater environments. This work aims to address and overcome these limitations.
 
\section{Methodology} \label{s3}

The underwater image enhancement involves the estimation of two main components: attenuation and backscatter.
The attenuation component is primarily controlled by the distance, while the backscatter component is affected by water type and ambient illumination. 
The OceanLens approach estimates these parameters using neural networks to accurately enhance underwater images.

\begin{figure}[h]
     \centering
     \includegraphics[width=0.85\linewidth]{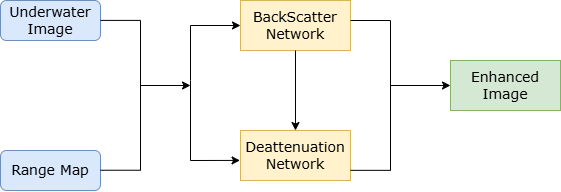}
     \caption{\small \textit{Illustration of the OceanLens Architecture.}}
     \label{nnarchflow}
 \end{figure}
The proposed OceanLens architecture operates with two key inputs: an underwater image and its corresponding range or depth map. These inputs are processed through two specialized neural networks—the backscatter correction network and the deattenuation network—working in tandem, as depicted in Fig. \ref{nnarchflow}. 

The backscatter correction network first analyzes the underwater image to estimate the backscatter component, which represents the scattered light that contributes to haze and reduced visibility. This estimated backscatter information is then passed on to the deattenuation network. The deattenuation network uses the estimated backscatter image to reconstruct the "direct image", representing the scene as it would appear without scattered light interference. Additionally, when combined with the range map, the direct image enables the network to compute the attenuation corrections required to restore the original underwater image. Subsequent sections provide more details on the proposed OceanLens architecture.

\subsection{Underwater Image Formation Method}
The underwater image captured by the camera is modeled as follows \cite{akkaynak2018}:
\begin{equation}
    I^c = I^c_D + I^c_{B} \label{eimgfor}
\end{equation}
where \( c \in \{R, G, B\} \) represents the color channel, \( I^c \) is the observed image distorted by the underwater environment, \( I^c_D \) is the \textit{"direct image"} including  attenuation effects, and \( I^c_{B} \) is the \textit{"backscatter"} resulting from reflections off suspended particles. The components \( I^c_D \) and \( I^c_{B} \) are given by:
\begin{eqnarray}
    I^c_D &=& I^c_J \exp(-a^c_D(z)z) \label{edirect} \\
    I^c_{B} &=& I^c_{B_{\infty}} (1 - \exp(-b^c z)) \label{ebs}
\end{eqnarray}
Here, \( z \) denotes the distance between the camera and the target object in the underwater scene. \( I^c_J \) represents the image of interest to be retrieved, and \( I_{B_{\infty}} \) is a parameter that quantifies the backscatter due to veiling light. The exponential terms in (\ref{edirect}) and (\ref{ebs}) are crucial as they model how signal attenuation and backscatter vary with range \( z \). Specifically, $a^c_D$ controls the attenuation in \( I^c_D \) and is a function of \( z \), while \( b^c \) governs the backscatter in \( I^c_{B} \).
Thus the $I^c_J$ can be reconstructed as,
\begin{equation}
    I^c_J = I^c_D(I^c_A)^{-1}, \label{eJ}
\end{equation}
where, $ I^c_D = I^c - I^c_B$ and 
\begin{equation}
    I^c_A = \exp(-a^c_D(z)z), \label{attcoff}
\end{equation}
where $a^c_D(z)$ is the coefficient of attenuation and it is a function of $z$. We provide details of how the OceanLens model estimates the backscatter component \( I^c_B \) and the attenuation component \( I^c_A \).
\subsection{Backscatter Modeling}
\begin{figure}
     \centering
     \includegraphics[width=0.85\linewidth]{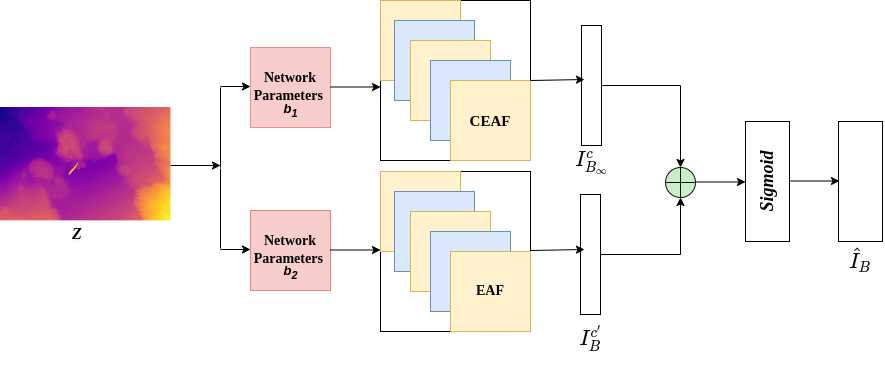}
     \caption{\small \textit{Network Architecture for Backscatter: The network takes Range map $Z$ as input and predicts backscatter image $\hat{I}_B$, with the kernel parameters in each convolution layer corresponding to those of the backscatter estimation model in \eqref{ebsnew}.}}
     \label{nnarch}
 \end{figure}

As we venture deeper into the ocean, the backscatter increases exponentially with depth \( z \) and eventually saturates. When all the light is absorbed or the image falls into complete darkness (shadow), the RGB intensity \( I^c \) approaches \( I^c_B \). The backscatter component \( I^c_B \), previously described in \eqref{ebs}, can be further modeled as:
\begin{equation}
    I^c_B = I^c_{B_{\infty}}(1 - \exp(-b^c_1z)) + I^{c'}_B\exp(-b^c_2z), \label{ebsnew}
\end{equation}
where the second term models a small residual component of the direct signal. Here, \( I^c_{B_{\infty}} \) and \( I^{c'}_B \) are network parameters. The BackscatterNet network is designed to estimate the direct component of an underwater image by isolating the backscatter effect. The values of these parameters typically range from \( 0 \) to \( 1 \). This is achieved through the neural network architecture depicted in Fig. \ref{nnarch}. First, the depth map \(D\) is processed through convolution layers to generate the initial coefficients: $b_1$ and $b_2$. The model consists of $P$ 2D-CNN layer.  Each term in \eqref{ebsnew} is implemented within the network as 

\begin{align}
    \hat{I}^c_B = \mathcal{F}^P_{p=1} [I^c_{B_{\infty}}(\mathcal{C}_{\mathcal{CEAF}}(-b_{p_1}Z))] +  \nonumber \\ \quad  \mathcal{F}^P_{p=1} [I^{c'}_B 
 (\mathcal{C}_{\mathcal{EAF}}(-b_{p_2}Z))], \label{ebsnewnn}
\end{align} 

Here, $P$ defines the number of convolutional layers, $\mathcal{F}$ represents concatenation, and $\mathcal{C}_x$ represents convolution with $x$ as its activation function. The Complementary Exponential Activation Function (CEAF) and Exponential Activation Function (EAF) are taken from \cite{Jamieson2023} and defined as.

\begin{equation}
  EAF(s) = \begin{cases}
1 &  s \leq 0 \\
\exp{(-s)} & s > 0 
\end{cases}  
\end{equation}
The outputs of the two networks, which correspond to two different terms, are summed and then passed through the sigmoid function, \( \sigma(s) = \frac{1}{1 + e^{-s}} \), to obtain the estimated backscatter \( \hat{I}^c_B \). Subsequently, the direct image component \( \hat{I}^c_D \) is estimated by subtracting \( \hat{I}^c_B \) from the original image \( I^c \), given by \( \hat{I}^c_D = I^c - \hat{I}^c_B \).

\subsubsection{Backscatter Loss Function}
We introduce an Adaptive Huber Loss function, given by \cite{sun2020}: 
\begin{equation}
  L_{B} = \begin{cases}
(\hat{I}_D)^2 & \text{if } | \hat{I}_D| \leq \delta \\
\beta \delta \cdot (|\hat{I}_D| -  \frac{\delta}{2}) & \text{otherwise},
\end{cases}  
\end{equation}
where, \( \delta \) is a threshold parameter controlling the transition between quadratic and linear loss. The constant $\beta$ is a hyperparameter that balances the relative importance of the two loss components. This adaptive loss is preferred because it combines robustness to outliers with smoothness and adaptability, offering a balance between positive and negative characteristics of $\hat{I}^c_D$. Also, helps in adapting to different image conditions, such as lighting variation.

\subsection{Deattenuation Modeling}

\begin{figure}
     \centering
     \includegraphics[width=0.85\linewidth]{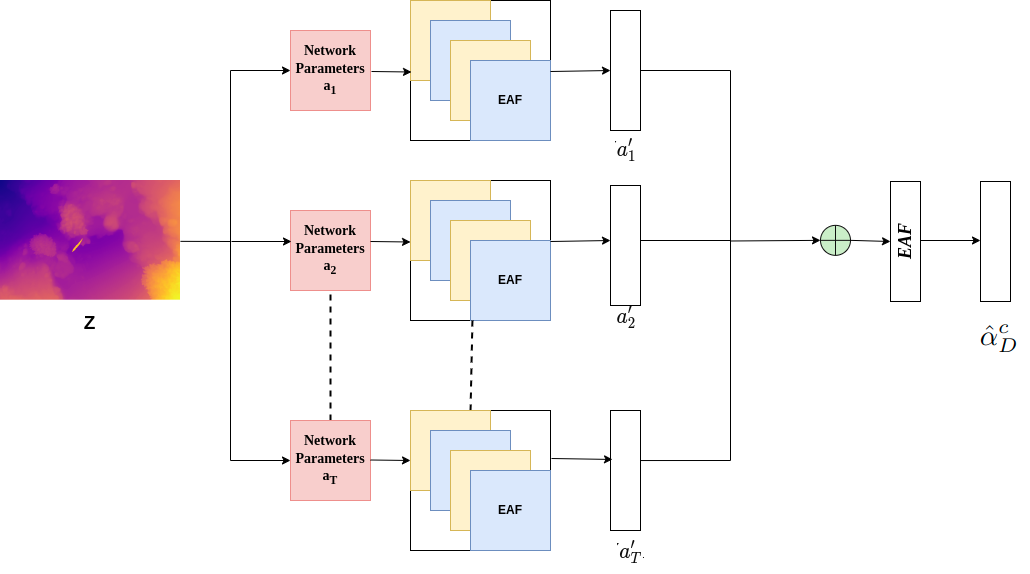}
     \caption{\small \textit{ Network Architecture for Deattenuation $\hat{\alpha}^c_D(Z)$: 
The network generates the $\hat{\alpha}^c_D(Z)$ from range map $Z$, with the kernel parameters $a^c_{p_i}$ in each convolution layer corresponding to the parameters of the attenuation coefficient $a^c_D$ function as described in \eqref{eatt}.}}
     \label{nnarchatt}
 \end{figure}

The deattenuation model aims to correct the observed image based on depth and attenuation factors. The coefficient of attenuation $a^c_D $ in \eqref{attcoff} is modeled as,
\begin{equation}
    a^c_D(z) = \sum_{p = 1}^{P}a'^c_p\exp(-a^c_pz), \label{eatt}
 \end{equation}
where, $a'_p, a_p \in [0~1]$,  $P$ represents the number of exponential functions that can approximate the attenuation. Thus the deattnuation $(I^c_A)^{-1}= \hat{\alpha}^c_D(Z) $ is estimated with neural network as,

\begin{align}
\hat{\alpha}^c_D(Z) = \mathcal{C}_{\mathcal{EAF}} \left( \mathcal{F}^P_{p_1=1} \left[ (Z * a_{p_1}^{lc}) \odot C_{\mathcal{EAF}}(Z * a_{p_1}^c) \right]+\right.\nonumber \\ \quad  \left., \ldots, +\mathcal{F}^P_{p_T=1} \left[ (Z * a_{p_T}^{lc}) \odot  \mathcal{C}_{\mathcal{EAF}}(Z * a_{p_T}^c) \right] \right),
\end{align}

where $T$ represents the number of exponentials and $\odot$ represents element-wise multiplication. The implementation of this in a similar way to a backscatter neural network as shown in Fig. \ref{nnarchatt}. However, we use only EAF as the activation function for the convolutional layers and the sigmoid function is replaced with EAF. The reconstructed image $\hat{I}^c_J$ is obtained as
\begin{equation}
\hat{I}^c_{J} = \hat{I}^c_D\hat{\alpha}^c_D(z)
\end{equation}

\subsubsection{Deattenuation Loss Function}
We employ a composite loss function that incorporates several components to capture various aspects of image quality. The overall loss function, \( L_{A} \), is defined as:
\begin{equation}
  L_{A} = L_{sat} + L_{int} + L_{var} + L_{sobel} + L_{log}.
\end{equation}
Each term in the loss function addresses the different attributes of the images. We use five different losses. The first loss function is the saturation loss \( L_{sat} \), which penalizes deviations of pixel values from the range \([0, 1]\). It is computed as:
\begin{align}
L_{sat} = \frac{1}{3} \sum_c \left( \frac{1}{N} \sum_{i=1}^{N} \left( \max(0, -\hat{I}_{J_i}) \right.\right. \nonumber \\ \quad \left. \left.  + \max(0, \hat{I}_{J_i} - I_{sat-tar}) \right)^2\right),
\end{align}
where subscript \(i \) denotes the pixel number of the estimate $\hat{I}_{J}$. The target saturation $I_{sat-tar}$ value of $1$ effectively enhances underwater images. 
It allows for flexible and localized adjustments, but care should be taken to avoid artifacts. 
Second, the intensity loss \( L_{int} \) enforces the desired target intensity for the image channels. It is calculated as:
\begin{equation}
 L_{int} = \frac{1}{3} \sum_c \left( \frac{1}{N} \sum_{i=1}^{N} (\bar{I}_{J_i} - I_{tar})^2\right),
\end{equation}
where \( \bar{I}_{J_i} \) is the mean intensity of the i and $I_{tar}$ is the target intensity values. It might be set to standard reference values such as mid-gray (e.g., 0.5 in normalized units) or white (e.g., 1.0). Third is variation loss \( L_{var} \), which measures the discrepancy in spatial variability between the input and reference images and is calculated as
\begin{equation}
L_{var} =  \frac{1}{3} \sum_c \left( SD(\hat{I}^c_{J}) - SD(\hat{I}^c_{D}) \right)^2,
\end{equation}
where \(SD(x)\) is the standard deviation (SD) of x. This tries to preserve the variance with the direct signal. 

Finally, we add two edge preservation losses to enhance the edges which play an important role in enhancement. Sobel Edge loss \( L_{sobel} \) evaluates the difference in edge information between the $ \hat{I}^c_{J}$ and direct signal $\hat{I}^c_{D}$. It is computed using the Sobel filters \( \mathbf{S_x} \) and \( \mathbf{S_y} \) to detect edges in both the $x$ and $y$ coordinates:
     \begin{align}
     L_{sobel} & = \frac{1}{3} \sum_{c} \left( \left| \mathbf{S_x} * \hat{I}^c_{J} - \mathbf{S_x} * \hat{I}^c_{D} \right| \right. \nonumber \\
     & \quad + \left. \left| \mathbf{S_y} * \hat{I}^c_{J} - \mathbf{S_y} * \hat{I}^c_{D} \right| \right),
\end{align}

 where \( * \) denotes convolution, \( \mathbf{S_x} \) and \( \mathbf{S_y} \) are the Sobel filters that compute the edge information and are defined as follows:
\begin{equation}
\mathbf{S_x} = \begin{bmatrix}
    1 & 0 & -1 \\
    2 & 0 & -2 \\
    1 & 0 & -1
\end{bmatrix}, \quad
\mathbf{S_y} = \begin{bmatrix}
    1 & 2 & 1 \\
    0 & 0 & 0 \\
    -1 & -2 & -1
\end{bmatrix}
\end{equation}
The next edge loss function introduced is the Laplacian of Gaussian loss  \( L_{log} \). This loss captures high-frequency variations by applying a Laplacian of Gaussian (LoG) filter, which is defined as follows:
\begin{equation}
    L_{log} = \frac{1}{3} \sum_{c}\left( \frac{1}{N}  \sum_{i=1}^N \left| \left( \mathbf{LoG}*(\hat{I}^c_{J_i}) \right) - \left( \mathbf{LoG}*(\hat{I}^c_{D_i}) \right) \right|\right),
 \end{equation}

where \( \mathbf{LoG}  \) represents the Laplacian of Gaussian filter.
The Gaussian kernel and Laplacian filter are defined as:
\begin{equation}
    \mathbf{G} = \frac{1}{16} \begin{bmatrix}
    1 & 2 & 1 \\
    2 & 4 & 2 \\
    1 & 2 & 1
    \end{bmatrix} , \quad
    \mathbf{L} = \begin{bmatrix}
    0 & -1 & 0 \\
    -1 & 4 & -1 \\
    0 & -1 & 0
    \end{bmatrix}
\end{equation}
Each component of the loss function is designed to address specific distortions and ensure that the output image closely matches the desired characteristics, resulting in an overall loss that balances saturation, intensity, spatial variation, edge preservation, and high-frequency details. 

\begin{figure*}
     \begin{center}
    \begin{tabular}{cccc}
    \includegraphics[width=0.18\linewidth, height=0.12\linewidth]{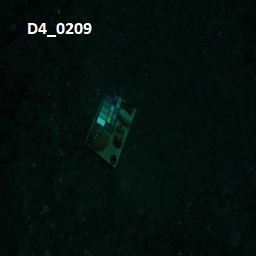}    & 
     \includegraphics[width=0.18\linewidth, height=0.12\linewidth]{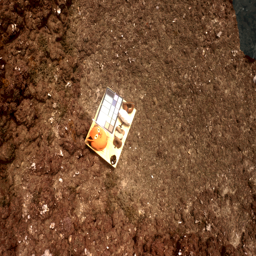} &  
       \includegraphics[width=0.18\linewidth, height=0.12\linewidth]{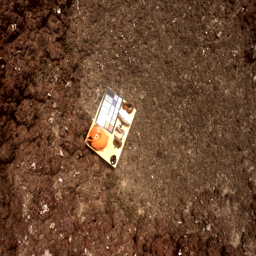} &  
       \includegraphics[width=0.18\linewidth, height=0.12\linewidth]{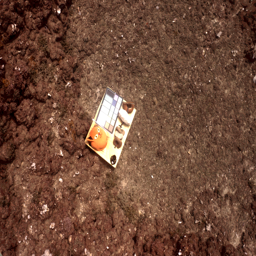} \\
       \includegraphics[width=0.18\linewidth, height=0.12\linewidth]{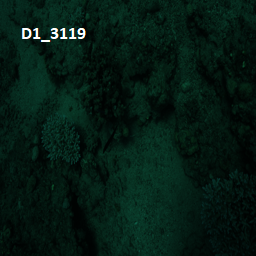}    & 
       \includegraphics[width=0.18\linewidth, height=0.12\linewidth]{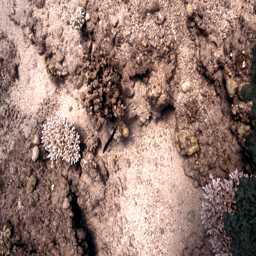} &  
       \includegraphics[width=0.18\linewidth, height=0.12\linewidth]{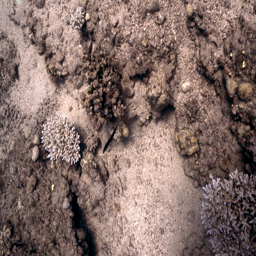} &  
       \includegraphics[width=0.18\linewidth, height=0.12\linewidth]{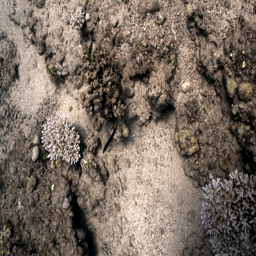} \\
       
       \includegraphics[width=0.18\linewidth, height=0.12\linewidth]{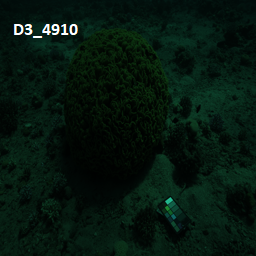}    &
       \includegraphics[width=0.18\linewidth, height=0.12\linewidth]{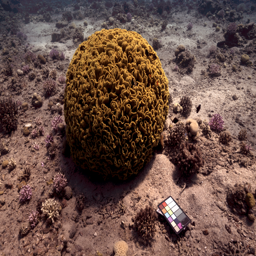} &  
       \includegraphics[width=0.18\linewidth, height=0.12\linewidth]{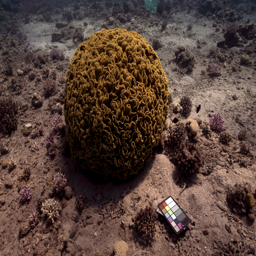} &  
       \includegraphics[width=0.18\linewidth, height=0.12\linewidth]{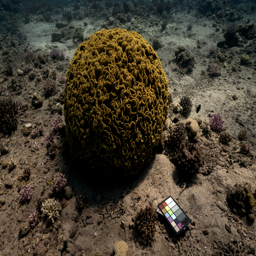} \\
       \includegraphics[width=0.18\linewidth, height=0.12\linewidth]{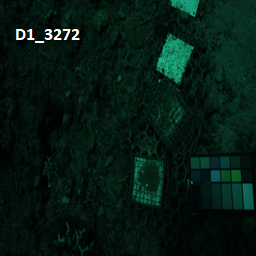}    &  
       \includegraphics[width=0.18\linewidth, height=0.12\linewidth]{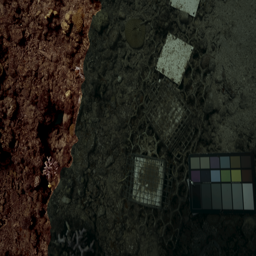} &  
       \includegraphics[width=0.18\linewidth, height=0.12\linewidth]{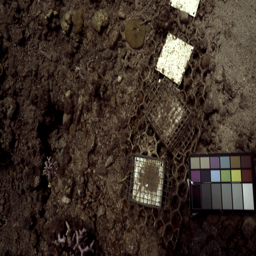} &
      \includegraphics[width=0.18\linewidth, height=0.12\linewidth]{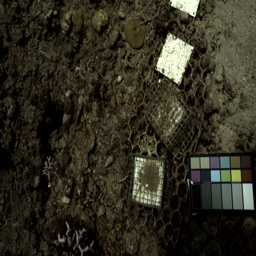} \\
       \includegraphics[width=0.18\linewidth, height=0.12\linewidth]{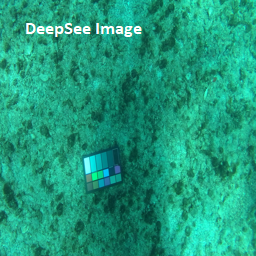}    &
       \includegraphics[width=0.18\linewidth, height=0.12\linewidth]{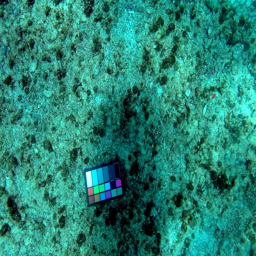} &  
       \includegraphics[width=0.18\linewidth, height=0.12\linewidth]{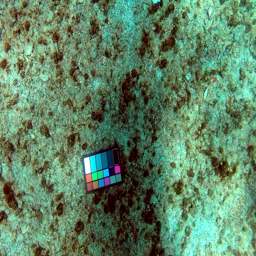} &  
       \includegraphics[width=0.18\linewidth, height=0.12\linewidth]{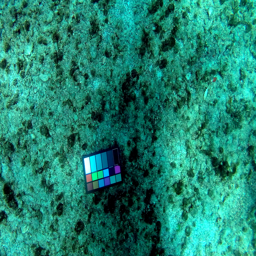} \\
       
       (a) & (b) & (c) & (d) 
    \end{tabular}
    \caption{\small \textit{(a) Raw Images from SeeThru \cite{akkaynak2018} (Row 1 - 4) and US Virgin Islands \cite{girdhar2023} dataset (Row 5). (b),~(c),~(d) Enhanced images obtained through OceanLens with one convolutional layer using ODM, MD and DA respectively.}}
    \label{CNN-single-layer}
    \end{center}
\end{figure*}

\section{Results}\label{s4}

\begin{table}[]
\renewcommand{\arraystretch}{1}
\setlength{\tabcolsep}{4pt}
\caption{\small \textit{ Gray Patch Mean Angular Error (GPMAE) values (in degrees) for images from SeeThru Dataset (D1 to D5) \cite{akkaynak2018} across various image enhancement methods: ST (SeeThru) \cite{akkaynak2019}, DSC (DeepseeColor) \cite{Jamieson2023}, and OceanLens (Ours) with three depth maps: Original Depth Map (ODM), MonoDepth2 (MD) \cite{godard2019}, and Depth-Anything-V2-Large (DA) \cite{yang2024v2}. Lower values indicate better performance, with \textbf{bold} values highlighting the best results.}}
\centering
\begin{tabular}{|c|c|c|c|ccc|}
\hline
\multirow{2}{*}{\textbf{Image}} & \multirow{2}{*}{\textbf{Raw}}                          & \multirow{2}{*}{\textbf{ST }} & \multirow{2}{*}{\textbf{DSC }} & \multicolumn{3}{c|}{\textbf{OceanLens}}\\ \cline{5-7} 
&&&& \multicolumn{1}{c|}{\textbf{ODM}}& \multicolumn{1}{c|}{\textbf{MD}}& \textbf{DA}\\ \hline
D1\_3272& 26& 8& 14& \multicolumn{1}{c|}{2.02}& \multicolumn{1}{c|}{2.61}& \textbf{1.2}\\ \hline
D2\_3647& 26& 8& 10& \multicolumn{1}{c|}{5.34}& \multicolumn{1}{c|}{3.63}& \textbf{1.22}\\ \hline
D3\_4910& 22& 8& 5& \multicolumn{1}{c|}{3.64}& \multicolumn{1}{c|}{2.97}&\textbf{0.57}\\ \hline
D4\_0209& 23& 4& 4& \multicolumn{1}{c|}{2.73}& \multicolumn{1}{c|}{1.19}&\textbf{1.04}\\ \hline
D5\_3374& \begin{tabular}[c]{@{}c@{}}17/16\\ /15/17\end{tabular} & \begin{tabular}[c]{@{}c@{}}\textbf{4/3}\\ \textbf{/5/3}\end{tabular}                & \begin{tabular}[c]{@{}c@{}}9/10\\ /10/11\end{tabular}              & \multicolumn{1}{c|}{\begin{tabular}[c]{@{}c@{}}11/28\\ /30/30\end{tabular}} & \multicolumn{1}{c|}{\begin{tabular}[c]{@{}c@{}}20/27\\ /29/28\end{tabular}} & \begin{tabular}[c]{@{}c@{}}9/20\\ /28/25\end{tabular} \\ \hline
\end{tabular}
    \label{tab:Ang_error_comparison}
\end{table}

\begin{table}[]
    \caption{\small \textit{ Underwater Image Quality Measure (UIQM) scores for OceanLens with one convolutional layer (Ours). Higher values indicate better performance, with \textbf{bold} values highlighting the best results.}}
    \renewcommand{\arraystretch}{1}
    \setlength{\tabcolsep}{2pt}
    \centering
    \begin{tabular}{|c|c|c|c|c|c|}
        \hline
        \multirow{2}{*}{\textbf{Image Name}} & \multirow{2}{*}{\textbf{Raw}} & \multirow{2}{*}{\textbf{Pre-processed}} & \multicolumn{3}{c|}{\textbf{OceanLens}} \\ \cline{4-6} 
                                &                           &                              & \textbf{ODM} & \textbf{MD} & \textbf{DA} \\ \hline
        D1\_3272 & 1.3748 & 1.9625 & \textbf{2.1523} & 1.8210 & 1.1451 \\ \hline
        D2\_3647 & 1.5029 & 2.4996 & 3.3734 & 2.9410 & \textbf{3.3916} \\ \hline
        D3\_4910 & 1.6124 & 1.4755 & \textbf{2.4545} & 2.4335 & 1.7935 \\ \hline
        D4\_0209 & 1.1404 & 2.0448 & 2.6454 & 2.6622 & \textbf{2.6628} \\ \hline
        D5\_3374 & 1.1648 & \textbf{1.7970} & 0.6348 & 0.6301 & 0.4452 \\ \hline
    \end{tabular}
    \label{tab:UIQMS_comparison}
\end{table}

\begin{table}[]
\caption{\small \textit{GPMAE values (in degrees) for images from SeeThru Dataset (D1 to D5) \cite{akkaynak2018}  for ST, DSC, and OceanLens with Multiple Convolutional layers (Ours).}}
\renewcommand{\arraystretch}{1}
\setlength{\tabcolsep}{4pt}
\centering
\begin{tabular}{|c|c|c|c|ccc|}
\hline
\multirow{2}{*}{\textbf{Image}} & \multirow{2}{*}{\textbf{Raw}}                          & \multirow{2}{*}{\textbf{ST}} & \multirow{2}{*}{\textbf{DSC}} & \multicolumn{3}{c|}{\textbf{OceanLens}}  \\ \cline{5-7}& & & & \multicolumn{1}{c|}{\textbf{ODM}} & \multicolumn{1}{c|}{\textbf{MD}} & \textbf{DA}\\ \hline
D1\_3272& 26& 8& 14& \multicolumn{1}{c|}{\textbf{2.25}}& \multicolumn{1}{c|}{9.72}&  4.33\\ \hline
D2\_3647& 26& 8& 10& \multicolumn{1}{c|}{\textbf{1.04}}& \multicolumn{1}{c|}{2.19}& 1.20\\ \hline
D3\_4910& 22& 8& 5& \multicolumn{1}{c|}{3.10}& \multicolumn{1}{c|}{2.50}& \textbf{0.40}\\ \hline
D4\_0209& 23& 4& 4& \multicolumn{1}{c|}{\textbf{2.50}}& \multicolumn{1}{c|}{3.05}& 4.81\\ \hline
D5\_3374& \begin{tabular}[c]{@{}c@{}}17/16\\ /15/17\end{tabular} & \begin{tabular}[c]{@{}c@{}}\textbf{4/3}\\ \textbf{/5/3}\end{tabular}                & \begin{tabular}[c]{@{}c@{}}9/10\\ /10/11\end{tabular}              & \multicolumn{1}{c|}{\begin{tabular}[c]{@{}c@{}}10/21\\ /32/9\end{tabular}} & \multicolumn{1}{c|}{\begin{tabular}[c]{@{}c@{}}19/27\\ 31/20\end{tabular}} & \begin{tabular}[c]{@{}c@{}}13/24\\ /31/14\end{tabular} \\ \hline

\end{tabular}
    \label{tab:Ang_error_comparison_multiple}
\end{table}

To demonstrate the efficacy of OceanLens, we evaluated underwater image enhancement using three datasets: See-Thru \cite{akkaynak2018}, US Virgin Islands \cite{girdhar2023}, and UIEB \cite{li2019}. These datasets, each with unique underwater conditions and image quality challenges, are ideal for testing our techniques. For depth maps, we utilized original maps and those generated by MonoDepth2 \cite{godard2019} and Depth-Anything-V2-Large \cite{yang2024,yang2024v2}, originally trained on terrestrial data. Our preprocessing involved white balancing and gamma correction. All models are implemented using PyTorch and are trained on an NVIDIA GeForce RTX 4090 GPU. On average, OceanLens takes approximately 4-5 milliseconds to process images with sizes ranging between 7 to 12 MB.

\begin{figure*}[h!]
    \centering
    \begin{tabular}{cccc}
    \includegraphics[width=0.18\linewidth, height=0.1\linewidth]{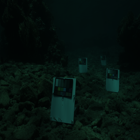} &
    \includegraphics[width=0.18\linewidth, height=0.1\linewidth]{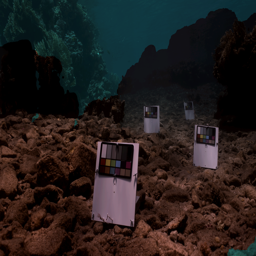} &
    \includegraphics[width=0.18\linewidth, height=0.1\linewidth]{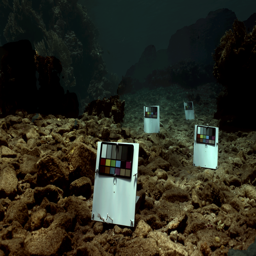} &
    \includegraphics[width=0.18\linewidth, height=0.1\linewidth]{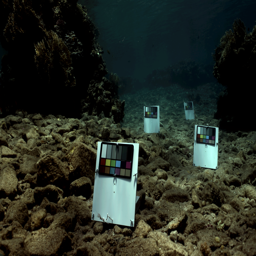} \\
     \includegraphics[width=0.18\linewidth, height=0.1\linewidth]{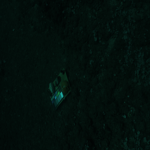} &
    \includegraphics[width=0.18\linewidth, height=0.1\linewidth]{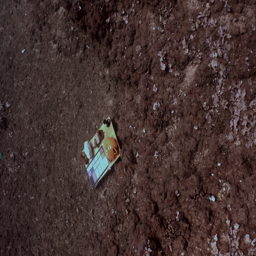} &
    \includegraphics[width=0.18\linewidth, height=0.1\linewidth]{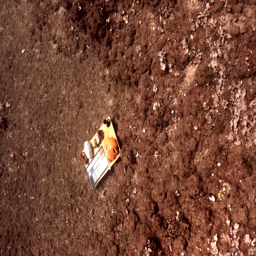} &
    \includegraphics[width=0.18\linewidth, height=0.1\linewidth]{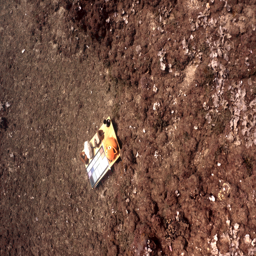} \\
     \includegraphics[width=0.18\linewidth, height=0.1\linewidth]{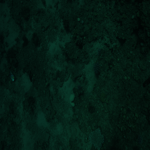} &
     \includegraphics[width=0.18\linewidth, height=0.1\linewidth]{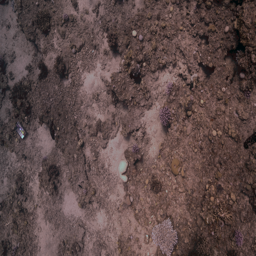} &
     \includegraphics[width=0.18\linewidth, height=0.1\linewidth]{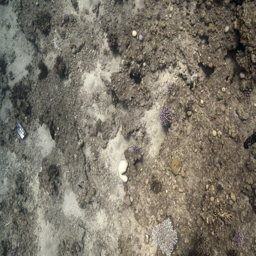} &
     \includegraphics[width=0.18\linewidth, height=0.1\linewidth]{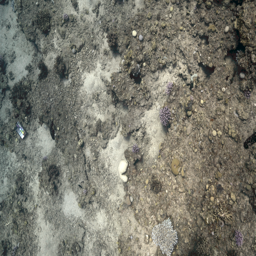} \\
     \includegraphics[width=0.18\linewidth, height=0.1\linewidth]{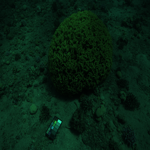} &
     \includegraphics[width=0.18\linewidth, height=0.1\linewidth]{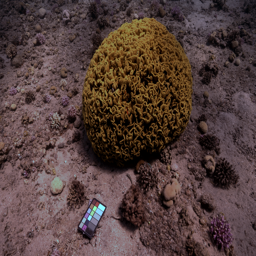} &
     \includegraphics[width=0.18\linewidth, height=0.1\linewidth]{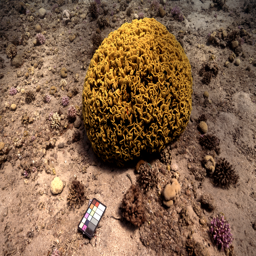} &
    \includegraphics[width=0.18\linewidth, height=0.1\linewidth]{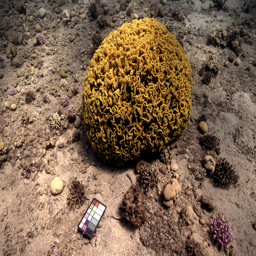} \\
    \includegraphics[width=0.18\linewidth, height=0.12\linewidth]{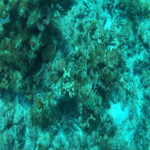} &
    \includegraphics[width=0.18\linewidth, height=0.12\linewidth]{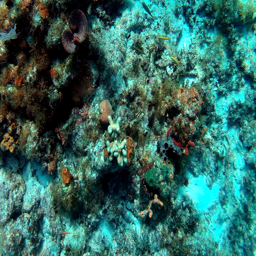}&
    \includegraphics[width=0.18\linewidth, height=0.12\linewidth]{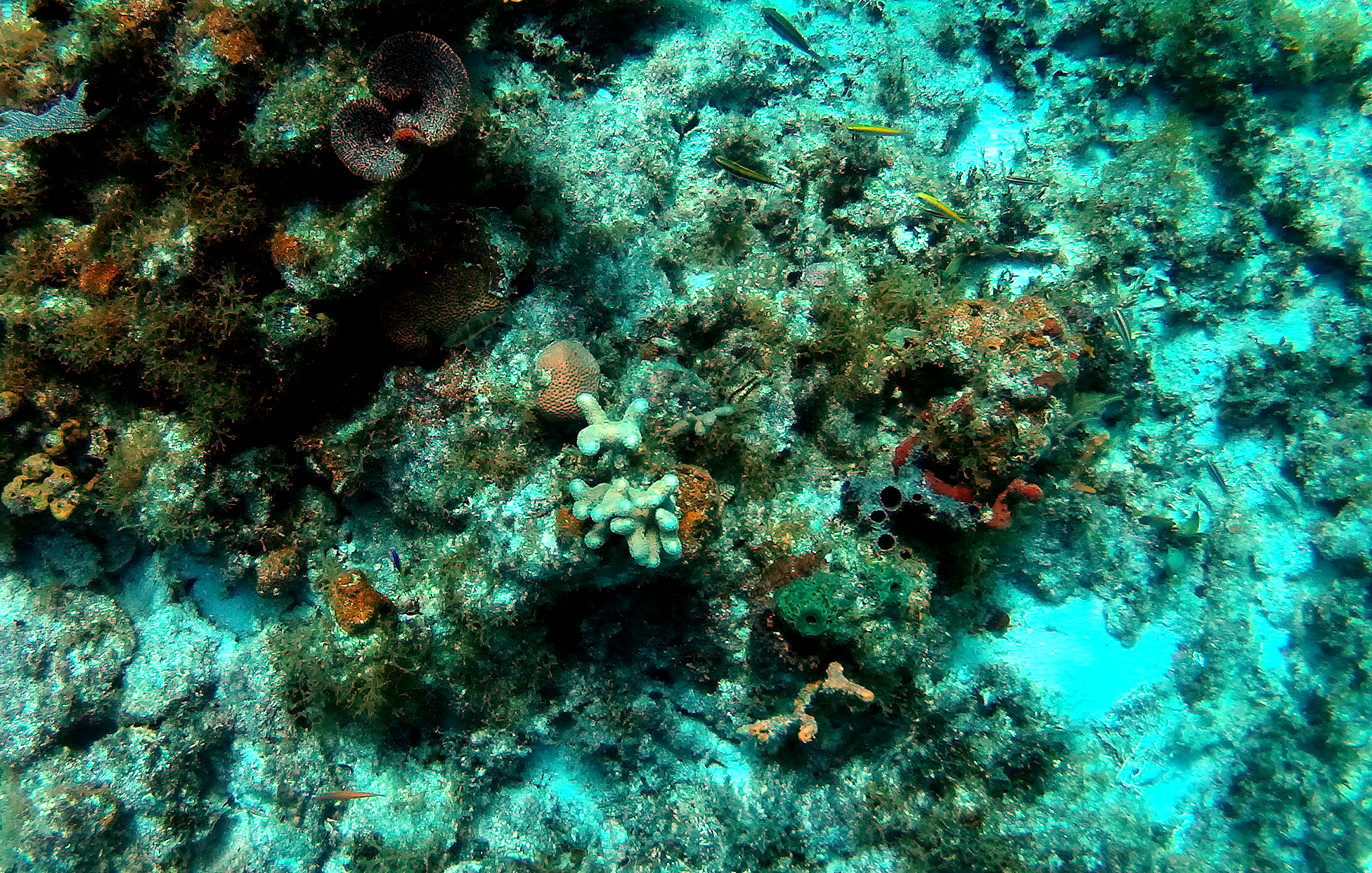}&
    \includegraphics[width=0.18\linewidth, height=0.12\linewidth]{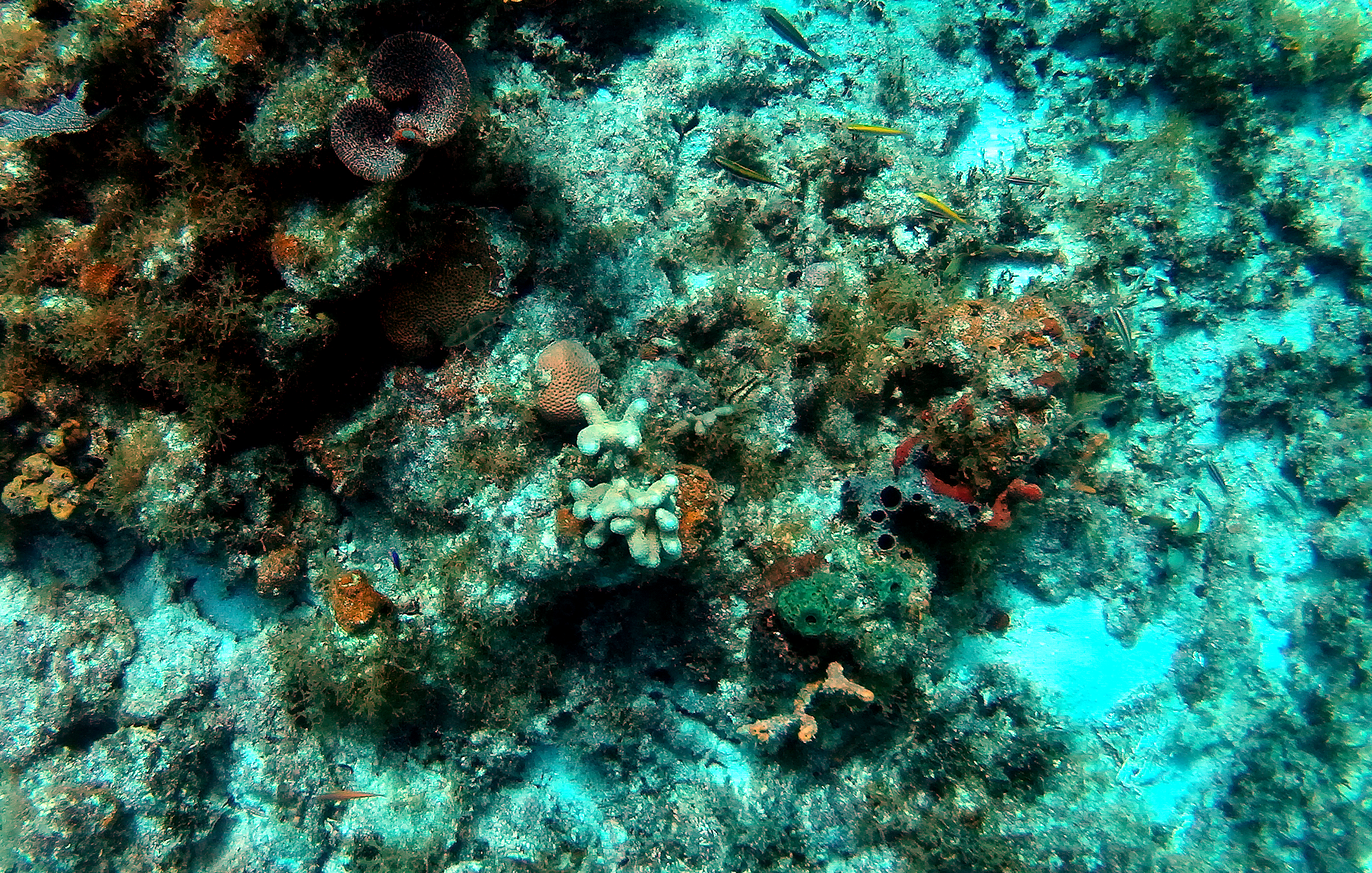}\\
    (a) & (b) & (c) & (d)
    \end{tabular}
    \caption{\small \textit{~(a) Raw Images from SeeThru (Row 1-4) and US Virgin Islands dataset (Row 5).~(b),~(c),~(d) Enhanced images obtained through OceanLens with multiple convolutional layer using ODM, MD and DA respectively.}}
    \label{2cnnDA}
\end{figure*}

\subsection{Evalution Metrics}
\subsubsection{Gray Patch Mean Angular error (GPMAE) \cite{berman2017}}
The average angular error is defined as

\begin{equation}
\bar{\psi} = \frac{1}{6} \sum_{x_i} \cos^{-1} \left( \frac{\mathbf{I}(x_i) \cdot (1, 1, 1)}{\|\mathbf{I}(x_i)\| \cdot \sqrt{3}} \right),
\end{equation}
where $x_i$ is the coordinates of the grayscale patches in the image. Lower angles indicate a more accurate color restoration.

\subsubsection{Underwater Image Quality Measure (UIQM) \cite{Yang2015}}
The overall underwater image quality measure is defined as
\begin{equation}
\text{UIQM} = c_1 \times \text{UICM} + c_2 \times \text{UISM} + c_3 \times \text{UIConM},
\end{equation}
where \(\text{UICM}\), \(\text{UISM}\), and \(\text{UIConM}\) represent the colorfulness, sharpness, and contrast measures, respectively, and are linearly combined to assess image quality and coefficients are \(c_1 = 0.0282\), \(c_2 = 0.2953\), and \(c_3 = 3.5753\). A greater UIQM value corresponds to an image with a better quality \cite{panetta2015}.

\begin{table}[]
    \caption{\small \textit{UIQM scores for OceanLens with multiple convolutional layers (Ours).}}
    \renewcommand{\arraystretch}{1}
    \setlength{\tabcolsep}{4pt}
    \centering
    \begin{tabular}{|c|c|c|c|c|c|}
        \hline
        \multirow{2}{*}{\textbf{Image Name}} & \multirow{2}{*}{\textbf{Raw}} & \multirow{2}{*}{\textbf{Pre-processed}} & \multicolumn{3}{c|}{\textbf{OceanLens}} \\ \cline{4-6} 
                                &                           &                              & \textbf{ODM} & \textbf{MD} & \textbf{DA} \\ \hline
        D1\_3272 & 1.3748 & 1.9625 & \textbf{2.1837} & 2.0226 & 1.5291 \\ \hline
        D2\_3647 & 1.5029 & 2.4996 & 3.3777 & 3.3741 & \textbf{3.4216} \\ \hline
        D3\_4910 & 1.6124 & 1.4755 & 1.9834 & 2.0252 & \textbf{2.1432} \\ \hline
        D4\_0209 & 1.1404 & 2.0448 & 2.4512 & 2.4952 & \textbf{2.7642} \\ \hline
        D5\_3374 & 1.1648 & \textbf{1.7970} & 0.7506 & 0.7054 & 0.7072 \\ \hline
    \end{tabular}
    \label{tab:UIQMS_comparison_multiple}
\end{table}

\subsubsection{Peak Signal-to-Noise Ratio (PSNR) and Structural Similarity Index (SSIM) \cite{panetta2015}}

PSNR measures image fidelity by comparing pixel values \cite{li2019}, with higher values indicating better noise reduction. SSIM  assesses perceptual image quality by evaluating structural details, luminance, and contrast. Increased SSIM signifies improved preservation of structural and textural details, reflecting better visual quality. Together, PSNR provides insight into noise suppression, while SSIM offers a more nuanced view of perceptual similarity and detail preservation.

\begin{figure}
    \begin{tabular}{ccc}
    \includegraphics[width=0.28\linewidth, height=0.18\linewidth]{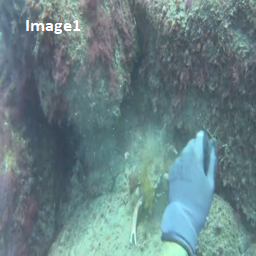} &
    \includegraphics[width=0.28\linewidth, height=0.18\linewidth]{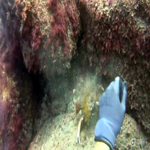} & 
    \includegraphics[width=0.28\linewidth, height=0.18\linewidth]{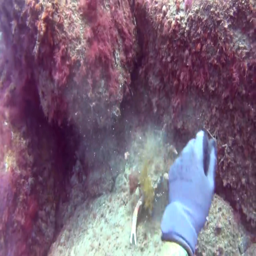} \\
    \includegraphics[width=0.28\linewidth, height=0.18\linewidth]{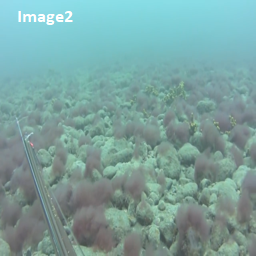} & 
     \includegraphics[width=0.28\linewidth, height=0.18\linewidth]{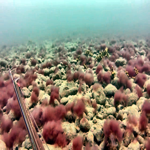} & 
    \includegraphics[width=0.28\linewidth, height=0.18\linewidth]{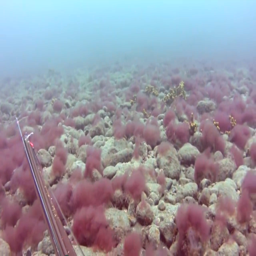} \\  \includegraphics[width=0.28\linewidth, height=0.18\linewidth]{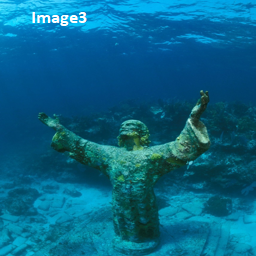} & 
    \includegraphics[width=0.28\linewidth, height=0.18\linewidth]{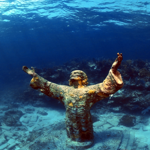} &
    \includegraphics[width=0.28\linewidth, height=0.18\linewidth]{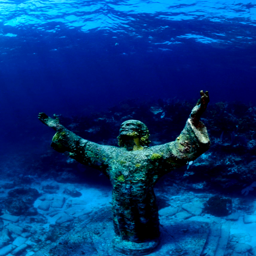} \\ 
    (a) & (b) & (c) 
    \end{tabular}
    \caption{\small \textit{(a) Raw Images from UIEB dataset \cite{li2019}. (b) Corresponding reference images. (c) Enhanced images - OceanLens with multiple convolutional layers using DA depth map.}}
    \label{UIEB}
\end{figure}

\begin{table}[]

    \caption{\small \textit{Comparison of the Peak Signal-to-Noise Ratio (PSNR) and the Structural Similarity Index Measure (SSIM) for OceanLens, utilizing multiple convolutional layers, against the UIEB benchmark dataset \cite{li2019} ($\uparrow$ : higher is better).}}
    \renewcommand{\arraystretch}{1}
    \setlength{\tabcolsep}{5pt}
    \centering
    \begin{tabular}{|c|c|c|c|c|}
        \hline
        \multirow{2}{*}{\textbf{Image}} & \multicolumn{2}{c|}{\textbf{Raw}} & \multicolumn{2}{c|}{\textbf{OceanLens}} \\ \cline{2-5} 
& \textbf{SSIM} ($\uparrow$) & \textbf{PSNR}  ($\uparrow$)   & \textbf{SSIM} ($\uparrow$) & \textbf{PSNR} ($\uparrow$)   \\ \hline
        Image1& 0.0428& 14.623& \textbf{0.413}   & \textbf{18.128}\\ \hline
        Image2& -0.0062& \textbf{8.520}& \textbf{0.0308}    & 7.773\\ \hline
        Image3& 0.678& \textbf{17.46}& \textbf{0.7209}  & 17.00\\ \hline
    \end{tabular}
    \label{tab:PSNR_SSIM_comparison}
\end{table}
 
\subsection{Experimental Results}

OceanLens demonstrates significant improvements in image quality over the SeeThru (ST) and DeepSeeColor (DSC) networks for a single convolutional layer as seen in Fig. \ref{CNN-single-layer}. Specifically, our method using Original Depth Map (ODM), MonoDepth2 (MD), and  Depth-Anything-V2-Large  (DA) respectively, achieves approximately 74.7\%, 67.3\%, and 85\% reductions in angular error for D1 as compared to ST, and 85.5\%, 81.3\%, and 91.42\% reduction as compared to DSC as seen in Table \ref{tab:Ang_error_comparison} . This trend is similar for D2, D3 and D4. Also, DA achieves the lowest angular error values, with 1.2 for D1 and 0.57 for D3. DA also shows notable reductions in GPMAE: a 40.6\% reduction for D1 compared to ODM and a 76.5\% reduction in D3 compared to MD. In terms of UIQM values, our method with ODM, MD, and DA shows improvements of 56.47\%, 
32.45\%, 5.5\% respectively, as compared to the raw image as seen in Table \ref{tab:UIQMS_comparison}. Among the three methods, DA generally outperforms ODM and MD.  For UIQM scores, we see an increasing trend with respect to all the depth maps. Our method, OceanLens, also excels at preserving the integrity of the color palette, particularly maintaining the vibrancy of colors such as yellow, brown, violet, and red.

While OceanLens performs well in many cases, it has notable failures. For instance, in Image D5-3374 (Table \ref{tab:Ang_error_comparison}) a low-lit underwater image, our approach results in a higher angular error than the ST and DSC methods. Moreover, the UIQM scores (Table \ref{tab:UIQMS_comparison}) for the raw D5 image are higher than those of the enhanced images, highlighting the need for pre-processing the low-lit underwater images. Despite using white balancing and gamma correction, the lower luminance in D5 contributed to this failure, emphasizing the difficulty of achieving optimal enhancement in all scenarios.

We also present results with OceanLens featuring multiple convolutional layers, tested on the datasets from \cite{akkaynak2019} and \cite{girdhar2023}.  Utilizing all the depth maps, the model performed well and we observe clear improvements in image enhancement compared to the raw images, as shown in Fig. \ref{2cnnDA}. UIQM scores and GPMAE are given in Table \ref{tab:UIQMS_comparison_multiple} and Table \ref{tab:Ang_error_comparison_multiple} respectively.  We emphasize that the model outperformed the one with a single convolutional layer.
Further validation was performed using the UIEB benchmark dataset \cite{li2019}, with network weights stored as checkpoints. The enhanced images obtained can be seen in Fig. \ref{UIEB}. Given the availability of reference images in UIEB, we evaluated our enhanced images using SSIM and PSNR metrics. Table \ref{tab:PSNR_SSIM_comparison} reveals that multi-layered convolutional in OceanLens consistently improved SSIM, reflecting better detail preservation compared to the raw images. For example, SSIM for \textit{Image3} increased by 6.3\%, and for \textit{Image2}, it rose from $-0.0062$ to $0.0308$. However, PSNR results showed mixed outcomes: while \textit{Image1} saw an increase (14.623 to 18.128), \textit{Image3} and \textit{Image2} experienced slight decreases (17.46 to 17.00 and 8.520 to 7.773, respectively), suggesting potential for further noise reduction improvements. These results demonstrate the effectiveness of our model in enhancing image structure while also indicating areas for further refinement.

\subsection{Ablation Study}
\begin{figure*}
     \centering
     \begin{tabular}{ccc}
         \includegraphics[width=0.3\linewidth, height=0.3\linewidth]{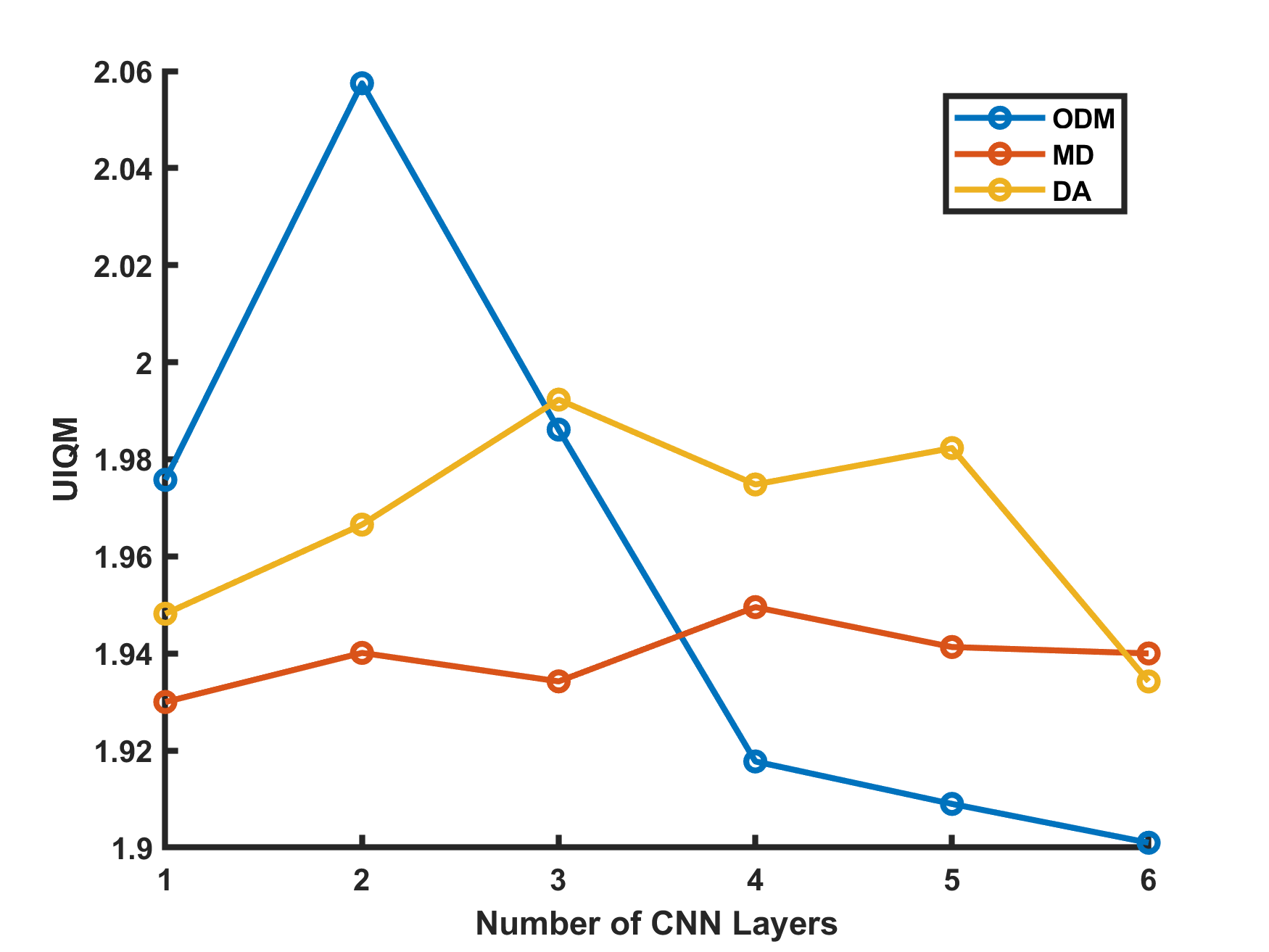} & \includegraphics[width=0.3\linewidth, height=0.3\linewidth]{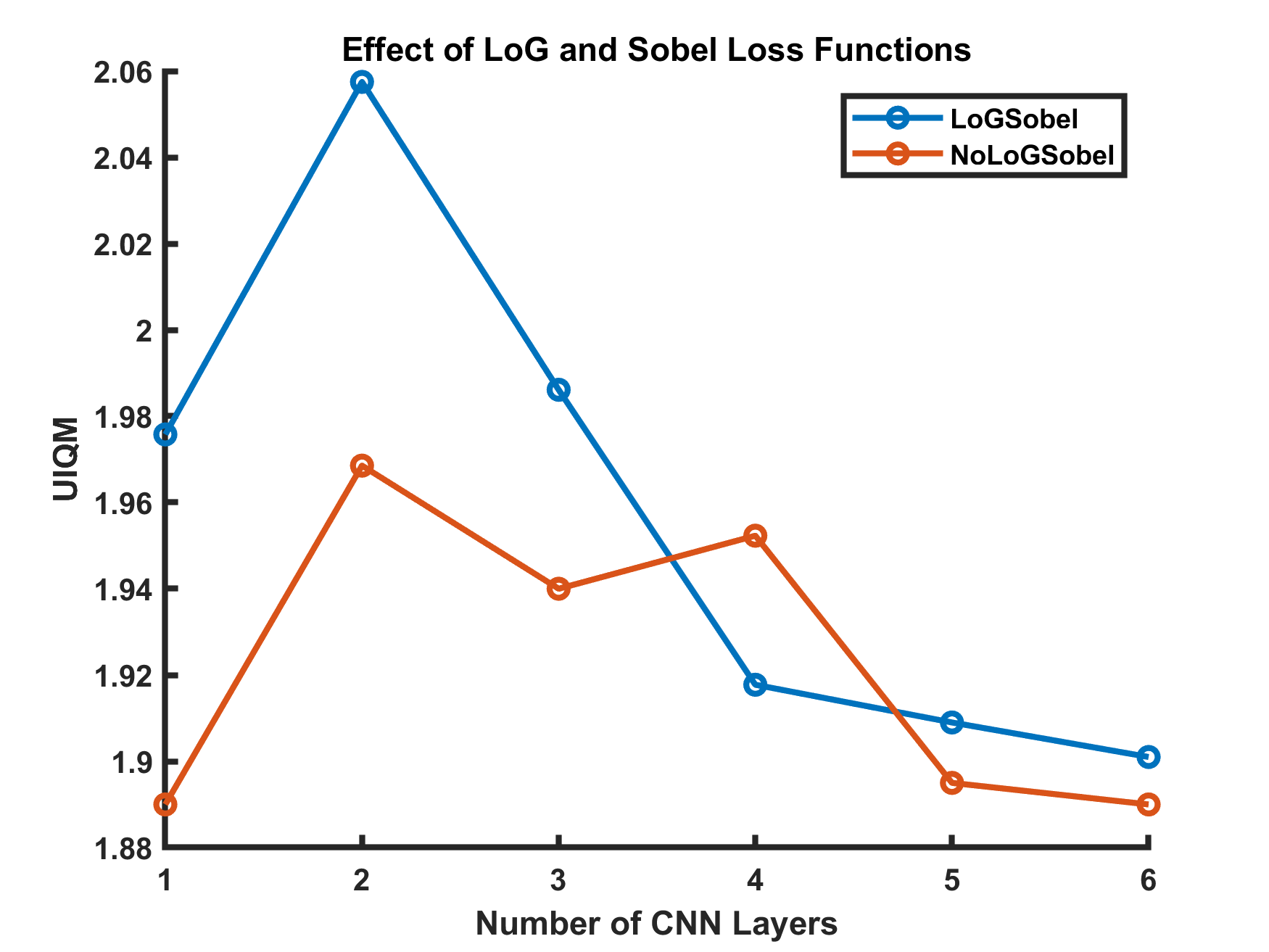} & \includegraphics[width=0.3\linewidth, height=0.3\linewidth]{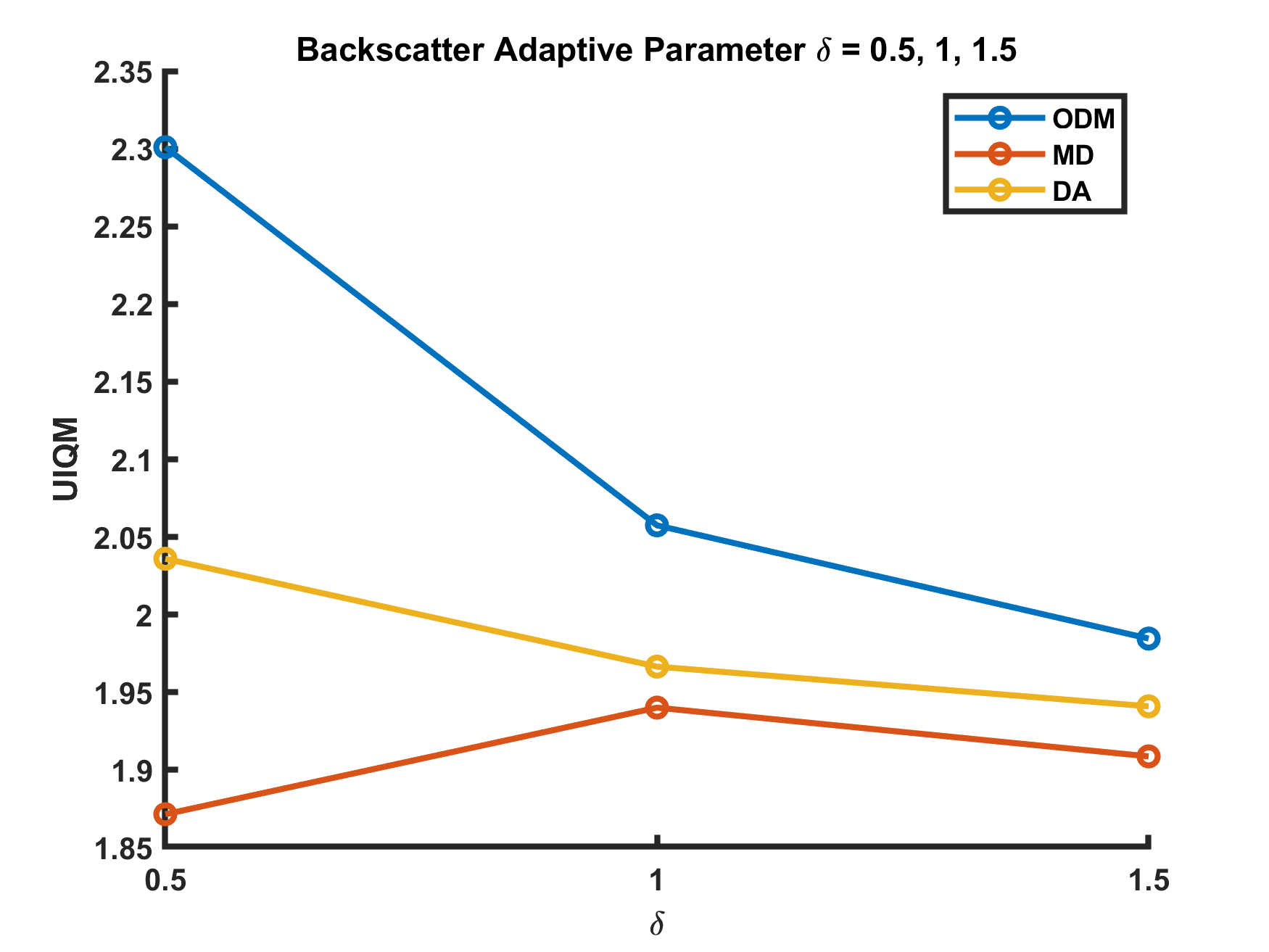}  \\
          (a) & (b)  & (c)
     \end{tabular}
     \caption{\small \textit{Variation of UIQM scores with multiple convolutional layers in OceanLeans: (a)~With different depth maps, (b)~With and without the use of Sobel and LoG loss functions and (c) Variation of UIQM scores with different $\delta$, an Adaptive Huber loss function parameter.}}
     \label{n_cnn_layer_graph}
 \end{figure*}

The results presented in the graphs illustrate the impact of various parameters on network performance. Fig. \ref{n_cnn_layer_graph} (a)  highlights how increasing the number of convolutional layers $N$ influences UIQM scores. For $N$ = 2 layers using ODM, $N$ = 3 layers using DA and $N$ = 4 layers using MD, the model demonstrates strong performance, as reflected in the increased UIQM scores. Fig. \ref{n_cnn_layer_graph} (b) demonstrates that the use of Sobel and LoG loss functions significantly enhances network performance. Fig. \ref{n_cnn_layer_graph} (c) shows the relationship between UIQM and $\delta$ indicating that an optimal value of $\delta$ = 0.5 yields superior performance.  Together, these findings emphasize the importance of parameter selection in optimizing performance for underwater image enhancement tasks.

\section{Conclusions}\label{s5}
In this paper, we introduced OceanLens, which offers a robust solution for tackling the challenges of underwater image enhancement by combining neural network architecture with specialized loss functions for edge preservation and accurate backscatter modeling. By integrating physics-based principles and utilizing monocular depth estimation models like MonoDepth2 and Depth-Anything-V2-Large, OceanLens effectively enhances image quality and fidelity. The method delivers substantial improvements, including a 65\% reduction in GPMAE and a 60\% boost in UIQM, showcasing its superiority over traditional methods such as SeeThru and DeepSeeColor. The performance of OceanLens on the UIEB benchmark dataset, with significant gains in SSIM, further demonstrates its ability to provide accurate corrections and facilitate near real-time processing. 

\balance


\begin{thebibliography}{10}

\bibitem{shuang2024}
X.~Shuang, J.~Zhang, and Y.~Tian, ``Algorithms for improving the quality of underwater optical images: A comprehensive review,'' {\em Signal Processing}, p.~109408, 2024.

\bibitem{akkaynak2017}
D.~Akkaynak, T.~Treibitz, T.~Shlesinger, Y.~Loya, R.~Tamir, and D.~Iluz, ``What is the space of attenuation coefficients in underwater computer vision?,'' in {\em Proceedings of the IEEE conference on computer vision and pattern recognition}, pp.~4931--4940, 2017.

\bibitem{Wang2019}
Y.~Wang, W.~Song, G.~Fortino, L.~Qi, W.~Zhang, and A.~Liotta, ``An experimental-based review of image enhancement and image restoration methods for underwater imaging,'' {\em IEEE Access}, vol.~7, pp.~140233--140251, 2019.

\bibitem{akkaynak2018}
D.~Akkaynak and T.~Treibitz, ``A revised underwater image formation model,'' in {\em Proceedings of the IEEE conference on computer vision and pattern recognition}, pp.~6723--6732, 2018.

\bibitem{jamieson2020}
S.~Jamieson, J.~P. How, and Y.~Girdhar, ``Active reward learning for co-robotic vision based exploration in bandwidth limited environments,'' in {\em 2020 IEEE International Conference on Robotics and Automation (ICRA)}, pp.~1806--1812, IEEE, 2020.

\bibitem{liang2023}
Z.~Liang, W.~Zhang, R.~Ruan, P.~Zhuang, X.~Xie, and C.~Li, ``Underwater image quality improvement via color, detail, and contrast restoration,'' {\em IEEE Transactions on Circuits and Systems for Video Technology}, 2023.

\bibitem{akkaynak2019}
D.~Akkaynak and T.~Treibitz, ``Sea-thru: A method for removing water from underwater images,'' in {\em Proceedings of the IEEE/CVF conference on computer vision and pattern recognition}, pp.~1682--1691, 2019.

\bibitem{zhu2020}
H.~Zhu, X.~Han, and Y.~Tao, ``Semi-supervised advancement of underwater visual quality,'' {\em Measurement Science and Technology}, vol.~32, no.~1, p.~015404, 2020.

\bibitem{Jamieson2023}
S.~Jamieson, J.~P. How, and Y.~Girdhar, ``Deepseecolor: Realtime adaptive color correction for autonomous underwater vehicles via deep learning methods,'' in {\em 2023 IEEE International Conference on Robotics and Automation (ICRA)}, pp.~3095--3101, 2023.

\bibitem{godard2019}
C.~Godard, O.~Mac~Aodha, M.~Firman, and G.~J. Brostow, ``Digging into self-supervised monocular depth estimation,'' in {\em Proceedings of the IEEE/CVF international conference on computer vision}, pp.~3828--3838, 2019.

\bibitem{yang2024v2}
L.~Yang, B.~Kang, Z.~Huang, Z.~Zhao, X.~Xu, J.~Feng, and H.~Zhao, ``Depth anything v2,'' {\em arXiv preprint arXiv:2406.09414}, 2024.

\bibitem{berman2017}
D.~Berman, T.~Treibitz, and S.~Avidan, ``Diving into haze-lines: Color restoration of underwater images,'' in {\em Proc. British Machine Vision Conference (BMVC)}, vol.~1, p.~2, 2017.

\bibitem{han2018}
M.~Han, Z.~Lyu, T.~Qiu, and M.~Xu, ``A review on intelligence dehazing and color restoration for underwater images,'' {\em IEEE Transactions on Systems, Man, and Cybernetics: Systems}, vol.~50, no.~5, pp.~1820--1832, 2018.

\bibitem{sanila2019}
K.~Sanila, A.~A. Balakrishnan, and M.~Supriya, ``Underwater image enhancement using white balance, usm and clhe,'' in {\em 2019 International Symposium on Ocean Technology (SYMPOL)}, pp.~106--116, IEEE, 2019.

\bibitem{hong2019}
N.~M. Hong and N.~Chi~Thanh, ``A single image dehazing method based on adaptive gamma correction,'' in {\em 2019 6th NAFOSTED Conference on Information and Computer Science (NICS)}, pp.~558--562, 2019.

\bibitem{hitam2013}
M.~S. Hitam, E.~A. Awalludin, W.~N. J. H.~W. Yussof, and Z.~Bachok, ``Mixture contrast limited adaptive histogram equalization for underwater image enhancement,'' in {\em 2013 International conference on computer applications technology (ICCAT)}, pp.~1--5, IEEE, 2013.

\bibitem{he2010}
K.~He, J.~Sun, and X.~Tang, ``Single image haze removal using dark channel prior,'' {\em IEEE transactions on pattern analysis and machine intelligence}, vol.~33, no.~12, pp.~2341--2353, 2010.

\bibitem{li2018}
C.~Li and X.~Zhang, ``Underwater image restoration based on improved background light estimation and automatic white balance,'' in {\em 2018 11th International Congress on Image and Signal Processing, BioMedical Engineering and Informatics (CISP-BMEI)}, pp.~1--5, IEEE, 2018.

\bibitem{zhou2014}
Q.-Y. Zhou and V.~Koltun, ``Color map optimization for 3d reconstruction with consumer depth cameras,'' {\em ACM Transactions on Graphics (ToG)}, vol.~33, no.~4, pp.~1--10, 2014.

\bibitem{de2021}
K.~De and M.~Pedersen, ``Impact of colour on robustness of deep neural networks,'' in {\em Proceedings of the IEEE/CVF international conference on computer vision}, pp.~21--30, 2021.

\bibitem{zhou2022}
J.~Zhou, D.~Zhang, and W.~Zhang, ``Underwater image enhancement method via multi-feature prior fusion,'' {\em Applied Intelligence}, vol.~52, no.~14, pp.~16435--16457, 2022.

\bibitem{an2024}
S.~An, L.~Xu, I.~Senior~Member, Z.~Deng, and H.~Zhang, ``Hfm: A hybrid fusion method for underwater image enhancement,'' {\em Engineering Applications of Artificial Intelligence}, vol.~127, p.~107219, 2024.

\bibitem{peng2022}
Y.-T. Peng, Y.-R. Chen, Z.~Chen, J.-H. Wang, and S.-C. Huang, ``Underwater image enhancement based on histogram-equalization approximation using physics-based dichromatic modeling,'' {\em Sensors}, vol.~22, no.~6, p.~2168, 2022.

\bibitem{chandrasekar2024}
A.~Chandrasekar, M.~Sreenivas, and S.~Biswas, ``Phish-net: Physics inspired system for high resolution underwater image enhancement,'' in {\em Proceedings of the IEEE/CVF Winter Conference on Applications of Computer Vision}, pp.~1506--1516, 2024.

\bibitem{Pham2023}
T.~T. Pham, T.~T.~N. Mai, and C.~Lee, ``Deep unfolding network with physics-based priors for underwater image enhancement,'' in {\em 2023 IEEE International Conference on Image Processing (ICIP)}, pp.~46--50, 2023.

\bibitem{li2019}
C.~Li, C.~Guo, W.~Ren, R.~Cong, J.~Hou, S.~Kwong, and D.~Tao, ``An underwater image enhancement benchmark dataset and beyond,'' {\em IEEE transactions on image processing}, vol.~29, pp.~4376--4389, 2019.

\bibitem{islam2020}
M.~J. Islam, Y.~Xia, and J.~Sattar, ``Fast underwater image enhancement for improved visual perception,'' {\em IEEE Robotics and Automation Letters}, vol.~5, no.~2, pp.~3227--3234, 2020.

\bibitem{li2017}
J.~Li, K.~A. Skinner, R.~M. Eustice, and M.~Johnson-Roberson, ``Watergan: Unsupervised generative network to enable real-time color correction of monocular underwater images,'' {\em IEEE Robotics and Automation letters}, vol.~3, no.~1, pp.~387--394, 2017.

\bibitem{yu2023}
B.~Yu, J.~Wu, and M.~J. Islam, ``Udepth: Fast monocular depth estimation for visually-guided underwater robots,'' in {\em 2023 IEEE International Conference on Robotics and Automation (ICRA)}, pp.~3116--3123, IEEE, 2023.

\bibitem{chen2023}
T.~Chen, N.~Wang, Y.~Chen, X.~Kong, Y.~Lin, H.~Zhao, and H.~R. Karimi, ``Semantic attention and relative scene depth-guided network for underwater image enhancement,'' {\em Engineering Applications of Artificial Intelligence}, vol.~123, p.~106532, 2023.

\bibitem{berman2020}
D.~Berman, D.~Levy, S.~Avidan, and T.~Treibitz, ``Underwater single image color restoration using haze-lines and a new quantitative dataset,'' {\em IEEE transactions on pattern analysis and machine intelligence}, vol.~43, no.~8, pp.~2822--2837, 2020.

\bibitem{bryson2016}
M.~Bryson, M.~Johnson-Roberson, O.~Pizarro, and S.~B. Williams, ``True color correction of autonomous underwater vehicle imagery,'' {\em Journal of Field Robotics}, vol.~33, no.~6, pp.~853--874, 2016.

\bibitem{hu2023}
K.~Hu, T.~Wang, C.~Shen, C.~Weng, F.~Zhou, M.~Xia, and L.~Weng, ``Overview of underwater 3d reconstruction technology based on optical images,'' {\em Journal of Marine Science and Engineering}, vol.~11, no.~5, p.~949, 2023.

\bibitem{liu2022}
D.~Liu, J.~Zhou, X.~Xie, Z.~Lin, and Y.~Lin, ``Underwater image restoration via background light estimation and depth map optimization,'' {\em Optics Express}, vol.~30, no.~16, pp.~29099--29116, 2022.

\bibitem{zhou2021}
J.~Zhou, T.~Yang, W.~Ren, D.~Zhang, and W.~Zhang, ``Underwater image restoration via depth map and illumination estimation based on a single image,'' {\em Optics Express}, vol.~29, no.~19, pp.~29864--29886, 2021.

\bibitem{sun2020}
Q.~Sun, W.-X. Zhou, and J.~Fan, ``Adaptive huber regression,'' {\em Journal of the American Statistical Association}, vol.~115, no.~529, pp.~254--265, 2020.

\bibitem{girdhar2023}
Y.~Girdhar, N.~McGuire, L.~Cai, S.~Jamieson, S.~McCammon, B.~Claus, J.~E. San~Soucie, J.~E. Todd, and T.~A. Mooney, ``Curee: A curious underwater robot for ecosystem exploration,'' in {\em 2023 IEEE International Conference on Robotics and Automation (ICRA)}, pp.~11411--11417, IEEE, 2023.

\bibitem{yang2024}
L.~Yang, B.~Kang, Z.~Huang, X.~Xu, J.~Feng, and H.~Zhao, ``Depth anything: Unleashing the power of large-scale unlabeled data,'' in {\em Proceedings of the IEEE/CVF Conference on Computer Vision and Pattern Recognition}, pp.~10371--10381, 2024.

\bibitem{Yang2015}
M.~Yang and A.~Sowmya, ``An underwater color image quality evaluation metric,'' {\em IEEE Transactions on Image Processing}, vol.~24, no.~12, pp.~6062--6071, 2015.

\bibitem{panetta2015}
K.~Panetta, C.~Gao, and S.~Agaian, ``Human-visual-system-inspired underwater image quality measures,'' {\em IEEE Journal of Oceanic Engineering}, vol.~41, no.~3, pp.~541--551, 2015.

\end{thebibliography}

\end{document}